\documentclass[%
 prd,
 twocolumn,
 superscriptaddress,
 numerical,
 showpacs,
 amsmath,amssymb,
 aps,
nofootinbib,
preprintnumbers,
longbibliography,
 floatfix
]{revtex4-2}
\usepackage[utf8]{inputenc}
\usepackage{amsmath}
\usepackage{amsfonts}
\usepackage{amssymb}
\usepackage{color}
\usepackage{graphicx}
\usepackage{epstopdf}
\usepackage{verbatim}
\usepackage{cancel}
\usepackage{siunitx}
\usepackage{braket}
\usepackage{bbold,ulem}
\usepackage[colorlinks=true, allcolors=blue]{hyperref}
\definecolor{DeepPink}{rgb}{0.796,0.004,0.384}
\definecolor{RoyalBlue}{rgb}{0.02,0.016,0.667}
\definecolor{Teja}{rgb}{0.65,0.20,0.15}
\definecolor{Berry}{rgb}{0.60,0.06,0.34}

\hypersetup{
    colorlinks=true,
    linkcolor=blue,
    urlcolor=cyan,
    citecolor=cyan,}

\usepackage{fancyvrb}
\usepackage{subfig}
\numberwithin{equation}{section}

\usepackage{multirow}
\usepackage{tikz}
 
\usepackage{array}

\begin{document}

\title{Small $x$ behavior in QCD from maximal entanglement and conformal invariance}

\author{Sebastian Grieninger}
\email{sebastian.grieninger@stonybrook.edu}
\affiliation{Center for Nuclear Theory, Department of Physics and Astronomy,
Stony Brook University, Stony Brook, New York 11794–3800, USA}
\affiliation{Co-design Center for Quantum Advantage (C2QA), Stony Brook University, New York 11794–3800, USA}
\author{Kun Hao}
\email{haoke72@163.com}
\affiliation{Peng Huanwu Center for Fundamental Theory, Xi’an 710127, China}
\affiliation{ Institute of Modern Physics, Northwest University, Xi’an 710127, China}
\author{Dmitri E. Kharzeev}
\email{dmitri.kharzeev@stonybrook.edu}
\affiliation{Center for Nuclear Theory, Department of Physics and Astronomy,
Stony Brook University, Stony Brook, New York 11794–3800, USA}
\affiliation{Co-design Center for Quantum Advantage (C2QA), Stony Brook University, New York 11794–3800, USA}
\affiliation{Energy and Photon Sciences Directorate, Condensed Matter and Materials Sciences Division,
Brookhaven National Laboratory, Upton, New York 11973-5000, USA}
\author{Vladimir Korepin}
\email{vladimir.korepin@stonybrook.edu}
\affiliation{Co-design Center for Quantum Advantage (C2QA), Stony Brook University, New York 11794–3800, USA}
\affiliation{C.N. Yang Institute for Theoretical Physics,
Stony Brook University, New York 11794, USA}
\date{\today}

\begin{abstract}
\noindent 
Recent evidence suggests that, at small Bjorken 
$x$, QCD evolution drives the proton into a state of maximal entanglement. If the evolution kernel is assumed to be conformally invariant -- as is the case
for the Balitsky-Fadin-Kuraev-Lipatov (BFKL) equation -- we can describe it by a conformal field theory. Moreover, the central charge 
$c$ of the corresponding conformal field theory emerges as the key parameter governing the 
$x$-dependence of both the entanglement entropy and the structure function.
Here we apply the exact Bethe Ansatz methods to the quantum spin chain dual to Lipatov's high energy effective action to extract the central charge of the theory, and find that $c=1$. This implies the $\sim x^{-1/3}$ small $x$ behavior for the structure function -- the prediction that can be tested at the forthcoming Electron-Ion Collider.

\end{abstract}

\maketitle

\section{Introduction}
Deep inelastic scattering (DIS) experiments at small Bjorken $x$ have revealed fascinating aspects of the strong force (see e.g. ~\cite{H1:1993jmo,Gribov:1983ivg,Mueller:1985wy,McLerran:1993ni,Badelek:1992gs,Gelis:2010nm,Nikolaev:1993th,ZEUS:1993ppj,Capella:1994cr,Ayala:1995hx,Jalilian-Marian:1997jhx,Frankfurt:1995jw,Ball:1995ye,Shigetani:1993dx,Nikolaev:1994cn,Abramowicz:1994gdc,Morfin:1990ck} and references therein).  
Understanding the gluon structure function at low $x$ is crucial to accurately describe DIS processes and to make predictions for physics experiments at the upcoming Electron-Ion Collider (EIC) at Brookhaven National Lab~\cite{Accardi:2012qut,AbdulKhalek:2021gbh,AbdulKhalek:2022hcn,Abir:2023fpo}. 
The key challenge lies in the resummation of so-called gluon ladders, $(\alpha_s\, \log(s))^n$, where $s$ is the squared center of mass energy and $\alpha_s$ the strong coupling constant, which spoil the convergence of the perturbative expansion even when the coupling $\alpha_s$ is small. 

In a series of seminal papers Balitsky, Fadin, Kuraev, and Lipatov (BFKL) developed an evolution equation that describes the high-energy behavior ($s\to\infty$) of QCD amplitudes by addressing the problem of the gluon ladder resummation~\cite{Fadin:1975cb,Balitsky:1978ic,Fadin:2020lam}.
Their approach, now known as BFKL theory, describes the scattering through reggeized gluon exchanges in the t-channel. The resulting evolution equations predict that gluon densities should grow as $x^{-\lambda}$ at small $x$, where the positive number $\lambda$ is the BFKL intercept computed in perturbation theory. Unfortunately, the inclusion of next-to-leading order corrections makes its value somewhat uncertain. Moreover, at high gluon density, the effects of gluon saturation are expected to set in, and modify the small $x$ behavior of the structure functions. In terms of BFKL theory, these effects correspond to multi-Pomeron interactions that are notoriously difficult to resum. 

In this paper, we propose to evaluate the small $x$ behavior of the gluon structure functions basing on the hypothesis of maximal entanglement put forward in \cite{Kharzeev:2017qzs}. If maximal entanglement is indeed reached at small $x$, the key quantity that determines the behavior of parton distributions is the entanglement entropy. If the effective QCD theory at small $x$ is conformally invariant, then the entanglement entropy depends solely on the central charge and Bjorken $x$ \cite{Kharzeev:2017qzs,Gursoy:2023hge}.

Let us briefly review the framework of Refs.~\cite{Kharzeev:2017qzs,Kharzeev:2021nzh} that establishes this connection. The proton in its rest frame is described by a pure quantum state $|\psi\rangle$ with density matrix $\hat{\rho} = |\psi\rangle\langle\psi|$ and vanishing von Neumann entropy $S = -\mathrm{tr}[\hat{\rho}\ln\hat{\rho}] = 0$. Deep inelastic scattering at Bjorken $x$ and virtuality $Q^2$ probes only a localized spatial region $A$ of the proton's wave function -- a tube of transverse size $\sim 1/Q$ and longitudinal extent $L \sim 1/(mx)$, where $m$ is the proton mass. The inclusive DIS cross section traces over the unobserved complementary region $B$, so the measurement accesses only the reduced density matrix $\hat{\rho}_A = \mathrm{tr}_B\,\hat{\rho}$. Consequently, DIS is characterized by an entanglement entropy
\begin{equation}
S_A = -\mathrm{tr}[\hat{\rho}_A \ln \hat{\rho}_A],\label{eq:SA}
\end{equation}
quantifying the entanglement between the probed region $A$ and the rest of the proton $B$.
\newline\newline
At small $x$, the probed region contains a large number of gluons, and the system is expected to approach a maximally entangled state~\cite{Kharzeev:2026jkq}. Physically, this corresponds to the regime of parton saturation, where gluon occupation numbers become large and the system explores all available microstates with equal probability -- an equipartitioned state. For $N$ equally probable microstates, the von Neumann entropy takes its maximal value $S = \ln N$.
Following Ref.~\cite{Kharzeev:2017qzs}, one further assumes that the effective number of accessible partonic microstates is proportional to the gluon distribution, $N\sim xG(x)$, which leads to
\begin{equation}
S(x) = \ln[xG(x)].
\label{eq:entropy_gluon}
\end{equation}
This relation provides the essential link between the entanglement entropy and the gluon distribution. As detailed in~\cite{Kharzeev:2017qzs}, Eq.~\eqref{eq:entropy_gluon} follows from Eq.~\eqref{eq:SA} after tracing over unobservable phases which is equivalent to tracing over lightcone time. This was shown using Mueller's dipole equation~\cite{Mueller:1994gb,Mueller:1993rr,Levin:2003nc}. Note that the underlying assumption in equating the two expressions is the maximal-entanglement picture introduced in~\cite{Kharzeev:2017qzs} (see \cite{Kharzeev:2026jkq} for further discussions).

If the effective theory governing small $x$ dynamics is conformally invariant, universal results from CFT constrain the entanglement entropy. For a $(1+1)$-dimensional CFT, the entanglement entropy of an interval of length $L$ with UV cutoff $\epsilon$ is given by the Calabrese-Cardy formula~\cite{Calabrese:2004eu,Calabrese:2005zw,Calabrese:2009qy}:
\begin{equation}
S_E = \frac{c}{3}\ln\frac{L}{\epsilon},
\label{eq:CFT_entropy}
\end{equation}
where $c$ is the central charge. In the DIS context, the longitudinal extent of the probed region provides the IR scale $L \sim 1/(mx) \sim 1/x$ (in units of the proton mass), while the UV cutoff $\epsilon$ is set by the proton's Compton wavelength $\sim 1/m$. Note that the UV cutoff is given by the Compton wavelength of the proton (and not by the frequency of the incoming photon) since we are in the rest frame of the proton. Substituting into Eq.~\eqref{eq:CFT_entropy}:
\begin{equation}
S(x) = \frac{c}{3}\ln\frac{1}{x}.
\end{equation}
Combining this with Eq.~\eqref{eq:entropy_gluon}, we obtain
\begin{equation}
\ln[xG(x)] = \frac{c}{3}\ln\frac{1}{x},
\end{equation}
which yields the small $x$ behavior of the gluon distribution:
\begin{equation}
xG(x) \sim x^{-c/3}.
\label{eq:xG_central_charge}
\end{equation}
Thus, the BFKL intercept $\lambda$ in the relation $xG(x) \sim x^{-\lambda}$ is directly determined by the central charge: $\lambda = c/3$. 

Hence, determining the central charge of the effective small $x$ theory is of direct phenomenological importance: it predicts the power-law growth of gluon structure functions. The assumption of maximally entangled state, combined with conformal invariance, reduces the problem of computing structure functions at small $x$ to identifying the CFT that governs the dynamics of reggeized gluons (in Lipatov's spin chain which is introduced in the following paragraph).

To carry out this program, we need an effective small $x$ QCD theory that allows to treat the exchange of an arbitrary number of reggeized gluons. Such an effective theory was constructed by Lipatov, who found that the equations governing reggeized gluon interactions in multicolor QCD can be reformulated as a quantum spin chain problem \cite{Lipatov:1993yb}. Specifically, the relevant Hamiltonian turned out to be equivalent to the integrable XXX model with zero spin.
This is not only satisfying from a mathematical stand point of view, but means that we can apply the powerful methods developed for exactly solvable models to high-energy QCD.
The spin chain picture becomes even richer when viewed through the lens of conformal field theory (CFT). In condensed matter physics, the connection between integrable chains and CFT has been a cornerstone for understanding critical phenomena.

Recent work has begun exploiting the spin chain picture to connect DIS with quantum information theory.  Based on the proposal of \cite{Kharzeev:2017qzs}, the authors of~\cite{Hao:2019cfu,Zhang:2021hra} studied the scattering process as a local quench in Lipatov's spin chain. They found that  the entanglement entropy in DIS grows logarithmically with time after the quench as $$S(t) = \frac{1}{3}\ln(t/\tau).$$
In their work, the authors implicitly assumed that the spin chain can be described by a CFT with central charge $c=1$. This leads to the behavior
$$xG(x) \sim x^{-1/3}$$ at small $x$ \cite{Kharzeev:2017qzs,Hentschinski:2021aux,Gursoy:2023hge,Hentschinski:2023izh,Hentschinski:2024gaa,Datta:2024hpn}.

The central charge $c$ is a fundamental invariant that characterizes conformal field theories and appears in many physical observables. In the specific context of integrable quantum systems studied via Bethe Ansatz methods, finite-size corrections provide the most direct path to extract the central charge.

The key advantage of the finite-size approach for integrable models is computational tractability. While the central charge appears in various CFT observables such as correlation functions, OPE coefficients, and anomaly terms, calculating these directly from the microscopic Hamiltonian of an interacting many-body system is typically intractable. In contrast, the Bethe Ansatz provides exact expressions for energy eigenvalues at finite size.

From the Virasoro algebra's representation theory, the finite-size energy of any CFT on a cylinder of circumference $L$ contains a universal $1/L$ correction:
\begin{equation}
E_{\text{finite}} = E_{\text{bulk}} - \frac{\pi c v_F}{6L} + \mathcal{O}(1/L^2),
\end{equation}
where the coefficient is completely determined by the central charge. For integrable models where we can compute $E_{\text{finite}}$ exactly via Bethe Ansatz, this formula provides a direct bridge between the microscopic quantum many-body problem and the universal properties of the emergent CFT.

Our work establishes that BFKL theory in the leading logarithmic approximation corresponds to $c = 1$ CFT. We reach this conclusion by employing the quantum inverse scattering method (QIS)~\cite{korepin1993quantuminversescatteringmethod} developed for integrable models. First, we exploit Faddeev and Korchemsky's mapping of the spin $0$ chain to a spin $-1$ chain model, which proves more tractable for Bethe Ansatz calculations. By computing finite-size corrections to the ground state energy, we determine the central charge unambiguously. Notably, our framework matches experimental data: the predicted Boltzmann-like relation between the gluon distribution and the entropy inferred from hadron multiplicity distributions has been tested at HERA at $x \sim 10^{-3}$--$10^{-4}$ and found to be in good agreement with data~(see \cite{Kharzeev:2026jkq} for a recent review).

These findings place BFKL theory within a broader theoretical framework connecting high-energy QCD, integrable systems, and quantum information. The identification of $c = 1$ constrains how structure functions behave at small $x$ and supports recent proposals about entanglement in parton distributions. Perhaps most intriguingly, our results suggest that quantum simulators designed for spin chains could provide new computational tools for studying high-energy scattering - a possibility that seemed remote just a few years ago. 

This paper is structured as follows. In section \ref{sec:framework}, we explain the connection between BFKL and the lattice nonlinear Schr\"odinger (NLS) equation. Then, in section \ref{sec:finitesize}, we compute the finite size corrections and extract the central charge. Finally, in section \ref{sec:conclusions}, we summarize our results and discuss future applications.
\section{Mapping BFKL to spin chains}\label{sec:framework}
The BFKL equation describing high-energy QCD scattering in the Regge limit is equivalent to an integrable quantum spin chain. The reggeized gluon Hamiltonian $H_L = \sum_{k=1}^L H_{k,k+1}$ with holomorphic coordinates exhibits $SU(2)$ symmetry and maps to the XXX Heisenberg model~\cite{Lipatov:1993yb,Faddeev:1994zg}. This establishes integrability through the Yang-Baxter equation, allowing exact solutions via Bethe Ansatz methods for computing high-energy QCD amplitudes. For more details see appendix~\ref{sec:framework2} and~\ref{sec:framework3} as well as \cite{Izergin:2009yc,Hao:2019cfu,Tarasov:1983cj} and references therein. The XXX model with negative spin $s = -1$ is equivalent to the quantum nonlinear Schr\"oedinger (NLS) equation, providing an alternative framework for studying the same physics.
\section{Finite Size Corrections}\label{sec:finitesize}
The energy of the ground state for BFKL model is (see for example appendix~\ref{sec:framework2})
\begin{equation}
    E=\sum_j\varepsilon_0(\lambda_j);\ \varepsilon_0(\lambda_j)=\frac{2}{\lambda_j^2+1}-h,
\end{equation}
where $h$ represents the chemical potential that controls the particle density in the grand canonical formulation. The ground state configuration corresponds to filling all single-particle states with negative energy $\varepsilon_0(\lambda_j) < 0$, which occurs for momenta satisfying $|\lambda_j| > \sqrt{2/h - 1}$ when $ 0 < h < 2$. 

In the thermodynamic limit $L \to \infty$, the discrete Bethe roots become continuously distributed, and the ground state energy density is determined by the solution $\varepsilon(\lambda)$ to the nonlinear integral equation
\begin{equation}
\varepsilon(\lambda) - \frac{1}{2\pi}\left( \int_{-\infty}^{-q}+ \int_{q}^{\infty}\right)K(\lambda, \mu)\varepsilon(\mu) d\mu = \frac{2}{\lambda^2 + 1} - h, \label{eq:energyintegral}
\end{equation}
subject to the conditions $\varepsilon(\pm q) = 0$. These conditions determine the Fermi surfaces (boundaries) $\pm q$ as functions of the chemical potential $h$, thereby establishing the relationship between the particle density and the thermodynamic parameters of the system.
The ground state configuration in the thermodynamic limit is characterized by a continuous distribution of Bethe roots. To describe this systematically, we introduce quantum numbers $n_j$ that label the allowed states. For a system with $N$ particles, the energy functional is minimized when these quantum numbers follow the pattern
\begin{equation}
    n_j = -\left(\frac{N-1}{2}\right) + j - 1, \quad j = 1,\ldots,N,
\end{equation}
corresponding to a filled Fermi sea configuration.

We define a smooth function $\lambda(x)$ that interpolates between the discrete Bethe roots $\{\lambda_j\}$ by establishing the relationship
\begin{equation}
    L\,\theta(\lambda(x))+\sum\limits_{j=1}^N\theta(\lambda(x)-\lambda_j)=2\pi\,x\,L\equiv 2\pi\,n_k.
\end{equation}
Here, $x$ serves as a continuous parameter, and $\lambda(n_j/L) = \lambda_j$ connects the discrete roots to the continuous description. In the case of NLS, the corresponding momentum function is given by
\begin{equation}
    \theta(\lambda)=p_0(\lambda)=i\,\log\,\left(\frac{i+\lambda}{i-\lambda}\right).
    \label{eq:bare-momentum}
\end{equation}

The thermodynamic limit $N, L \to \infty$ with fixed density $D = N/L$ transforms the discrete distribution of Bethe roots into a continuous density. In this regime, we can think of the momentum space as being populated by "particles" (occupied states) and "holes" (unoccupied states), with the totality of both constituting "vacancies." The density of vacancies in momentum space is defined as
\begin{equation}
    \rho_t(\lambda(x))=\frac{\text{d}x(\lambda)}{\text{d}\lambda},
\end{equation}
which quantifies how the continuous parameter $x$ varies with respect to the momentum $\lambda$. This density satisfies the integral equation
\begin{equation}
    2\pi \rho_t(\lambda)=\frac{1}{L}\sum\limits_{j=1}^N K(\lambda(x),\lambda_j)+K(\lambda(x)),\label{eq:dens}
\end{equation}
where $K(\lambda,\mu)=\frac{2\kappa}{\kappa^2+(\lambda-\mu)^2}$ is the interaction kernel and $K(\lambda)=K(\lambda,0)$ represents the boundary contribution. We focus on the case $\Delta=2$ and $\kappa=1$, for which the spin takes the value $s=-1=-\frac{2}{\kappa \Delta}$. Here $\Delta$ is the lattice spacing and $\kappa$ is the coupling constant of the lattice nonlinear Schr\"odinger equation, (see appendix \ref{sec:framework3}).
With these parameters, one immediately finds that $\theta'(\lambda)=p_0'(\lambda)=K(\lambda)$,
where $\theta(\lambda)$ is the scattering phase and $p_0(\lambda)$ the bare momentum \eqref{eq:bare-momentum}.

In the ground state, all single-particle states with negative energy are occupied, forming what is analogous to a Fermi sphere in momentum space. In the energy representation \eqref{eq:energyintegral}, the occupied region consists of the intervals $(-\infty,-q]\cup[q,\infty)$, where $q$ is the Fermi surface (rapidity) determined by the chemical potential. The quantity $L\rho(\lambda)d\lambda$ gives the number of particles (occupied states) in the momentum interval $[\lambda, \lambda + d\lambda]$, providing a direct connection between the Bethe Ansatz description and the macroscopic thermodynamic properties. 

The Fermi surface $q = \lim_{N\to\infty} \lambda_N$ marks the edge of the filled region, and in the thermodynamic limit, the discrete sums over Bethe roots can be replaced by continuous integrals $\sum_j \to L {\left(\int_{-\infty}^{-q}+\int^{\infty}_{q}\right)} \rho(\lambda) d\lambda$, since the spacing between adjacent roots becomes infinitesimal. 

However, for finite system sizes, systematic deviations from this continuum approximation arise, leading to finite-size corrections that are essential for understanding the conformal properties of the underlying field theory.

To capture these finite-size effects systematically, we employ the Euler-Maclaurin formula, which provides a precise expansion for approximating sums by integrals. The leading finite-size correction to the density equation transforms our discrete sum into~\cite{abramowitz1964}:

\begin{align}
   & \rho_L(\lambda)-\frac{1}{2\pi}{\left(\int_{-\infty}^{-q}+\int^{\infty}_{q}\right)}K(\lambda,\mu)\rho_L(\mu)\,\text{d}\mu\label{eq:rhoapprox}\\&=\frac{1}{2\pi}\left(p_0'(\lambda)+\frac{1}{24L^2\rho(q)}[K'(\lambda,-q)-K'(\lambda,q)]\right),\nonumber
\end{align}

where $\rho_L(\lambda)$ denotes the finite-size density that differs from the thermodynamic limit density $\rho(\lambda)$. The right-hand side now contains the finite-size correction term proportional to $1/L^2$, which arises from the discrete nature of the Bethe roots and the boundary effects at the Fermi surface $\pm q$.

The derivation involves a change of variables from the discrete index $j$ to the continuous momentum $\lambda$. Since the density relates these variables through $d\mu = dy/\rho_t(\mu)$, where $y$ is the continuous coordinate parameter,\footnote{Here, we used that $f(\lambda(y))dy=f(\mu)$, with \newline $\mu=\lambda(y)\leftrightarrow d\mu=(d\lambda/dy(\lambda))\, dy(\lambda)\leftrightarrow d\mu= dy/\rho$.} the transformation of derivatives follows $\partial_y \cdot = (1/\rho_t) \partial_\mu \cdot$.

Applying the same Euler-Maclaurin expansion to the total energy yields:
\begin{equation}
\!\!\!    E=L {\left(\int_{-\infty}^{-q}+\int^{\infty}_{q}\right)}\varepsilon_0(\lambda)\rho_L(\lambda)\,\text{d}\lambda-\frac{\pi}{6L}\frac{\varepsilon'(q)}{2\pi\rho(q)},\label{eq:epsapprox}
\end{equation}

where the finite-size correction exploits the antisymmetry $\varepsilon'(q) = -\varepsilon'(-q)$ of the derivative of the single-particle energy about the origin. 

A subtle but important point concerns the integration limits. While the natural upper limit is $q = \lambda(x = N/(2L))$, the actual uppermost Bethe root corresponds to $\lambda_N = \lambda(x = (N-1)/(2L))$. This discrepancy introduces an additional correction $\Delta\lambda = \lambda(x=N/(2L)) - \lambda_N = (d\lambda/dx) \cdot (1/(2L)) = 1/(2L\rho_t)$, which contributes to the finite-size scaling and must be accounted for. In this equation we used that $\Delta x=1/(2L).$
Hence, the precise relationship between the Fermi momentum and the uppermost Bethe root is:
$    q\equiv\lambda_N+\frac{1}{2L\rho}.$

The finite-size density equation~\eqref{eq:rhoapprox} can be solved formally using the resolvent method. Treating the integral operator $(1-\frac{1}{2\pi} K)^{-1}$ as an inverse, we obtain:

\begin{align}
    \rho_L=&\frac{1}{2\pi}\left(\int_{-\infty}^{-q}+\int^{\infty}_{q}\right) d\mu \left(1-\frac{1}{2\pi} K\right)^{-1}\!\!\!\!\!(\lambda,\mu)\nonumber\\&\times\left[p'_0(\mu)+\frac{1}{24L^2\rho(q)}(K'(\mu,-q)-K'(\mu,q))\right].\label{eq:rhosol}
\end{align}

Similarly, we can solve the defining equation for $\varepsilon(\mu)$ \eqref{eq:energyintegral} formally and establish:
\begin{equation}
    {\left(\int_{-\infty}^{-q}+\int^{\infty}_{q}\right)}d\lambda\, \varepsilon_0(\lambda)\left(1-\frac{1}{2\pi} K\right)^{-1}\!\!\!(\lambda,\mu)=\varepsilon(\mu).\label{eq:invk}
\end{equation}

Substituting the formal solution~\eqref{eq:rhosol} into the energy expression~\eqref{eq:epsapprox} and applying the identity~\eqref{eq:invk}, we obtain (keeping the integration limits in mind):
\begin{align}
    &E=\frac{L}{2\pi}\int d\mu\,\varepsilon(\mu)p'_0(\mu)-\frac{\pi}{6L}\frac{\epsilon'_0(q)}{2\pi\rho(q)}+\frac{1}{48\pi\,L\,\rho}\times\nonumber\\&\times\int\!\! d\mu\,\int\!\! d\lambda\,\varepsilon_0(\lambda)\left(1-\frac{1}{2\pi} K\right)^{-1}\!\!(K'(q,-\mu)-K'(q,\mu)).\label{eq:enew}
\end{align}

This expression can be simplified further by differentiating the integral equation for $\varepsilon(q)$~\eqref{eq:energyintegral} with respect to the Fermi surface (boundary) parameter $q$. Since $\varepsilon(\pm q) = 0$ by definition, the boundary terms vanish upon differentiation, yielding:

\begin{widetext}
 \begin{equation}
    \varepsilon'(q)=\varepsilon_0'(q)+\frac{1}{2\pi}{\left(\int_{-\infty}^{-q}+\int^{\infty}_{q}\right)}K'(q,\mu)\varepsilon(\mu) d\mu=\varepsilon_0'(q)+\frac{1}{4\pi}{\left(\int_{-\infty}^{-q}+\int^{\infty}_{q}\right)}(K'(q,-\mu)-K'(q,\mu))\varepsilon(\mu) d\mu,\label{eq:epsder}
\end{equation}   
\end{widetext}

where the second equality exploits the antisymmetry of the integrand under $\mu \to -\mu$.

Rearranging equation~\eqref{eq:epsder} to isolate $\varepsilon_0'(q)$ and substituting back into~\eqref{eq:enew}, the terms proportional to the kernel cancel, leaving us with:
\begin{equation}
     E=\frac{L}{2\pi}{\left(\int_{-\infty}^{-q}+\int^{\infty}_{q}\right)} d\mu\,\varepsilon(\mu)p'_0(\mu)-\frac{\pi}{6L}v_F,
\end{equation}
where $v_F=\frac{\varepsilon'(q)}{2\pi\rho(q)}$ is the Fermi velocity, representing the group velocity of excitations at the Fermi boundary.

This energy expression has the characteristic form predicted by conformal field theory for a critical system on a finite-size torus. According to the Virasoro algebra's representation theory, the finite-size energy of a conformal field theory takes the universal form:
\begin{equation}
     E_{\text{CFT}}=E_{\text{bulk}}-c\frac{\pi v_F}{6L},
\end{equation}
where $c$ is the central charge of the corresponding Virasoro algebra, which characterizes the conformal anomaly and determines the finite-size scaling behavior.

Comparing our exact result with the CFT prediction, we immediately extract:
$$c=1.$$

This result establishes that the quantum lattice nonlinear Schr\"odinger model, and hence the underlying BFKL equation through the established correspondence, belongs to the universality class of conformal field theories with central charge $c=1$. This places the high-energy QCD dynamics described by BFKL in the same conformal class as free fermions and compactified bosons, revealing deep connections between gauge theory physics and exactly solvable statistical mechanical models.

\section{Conclusions and Outlook}
\label{sec:conclusions}

In this work, we described small $x$ QCD as a conformal field theory, building on Lipatov's spin chain duality. Using exact Bethe Ansatz methods, we extracted the central charge of the corresponding CFT and found $c=1$. This result provides a solid theoretical foundation for earlier proposals linking parton distributions to quantum entanglement, and implies the universal small $x$ behavior 
\begin{equation}
xG(x) \sim x^{-1/3}.
\end{equation}
The identification of $c=1$ places the BFKL dynamics within a well-understood universality class and opens the door to applying powerful CFT techniques to high-energy QCD, where conventional methods are notoriously difficult.  
The CFT framework allows for systematic generalizations. In particular, corrections to the gluon structure function at small Bjorken $x$ can be computed using perturbed CFT techniques. We anticipate an expansion of the form  
\begin{equation}
    xG(x) = \sum_{n=-1}^\infty a_n \, x^{n/3},
\end{equation}
where the leading $n=-1$ term was obtained in~\cite{Zhang:2021hra}, while the higher-order coefficients $a_n$ remain to be determined. Their calculation will require extending the methods developed here to incorporate perturbations of the $c=1$ theory. 

It will be interesting to confront the predicted $x^{-1/3}$ scaling at small $x$ with future Electron-Ion Collider (EIC) data. Precise measurements of structure functions in the low-$x$ regime may help assess the validity of the entanglement-based framework considered here.
Note, however, that our framework does not include any effects of the running coupling since this would break conformal invariance. Hence, we cannot address the $Q^2$ dependence of the intercept which is of phenomenological importance and our analysis is restricted to a representative fixed large value of $Q^2$. Finally, the extraction of deviations from the leading $x^{-1/3}$ behavior at the EIC may provide experimental access to the higher-order coefficients $a_n$, thereby offering a direct probe of perturbed conformal dynamics in QCD.

\section*{Acknowledgments}
We thank Amanda Cooper-Sarkar for correspondence about~\cite{Abt:2017nkc}.
This work was supported by the U.S. Department of Energy, Office of Science, Office of Nuclear Physics, Grants No. DE-FG88ER41450
(D.K.) and DE-SC0012704 (S.G., D.K., V.K.) and
by the U.S. Department of Energy, Office of Science, National Quantum Information Science Research Centers,
Co-design Center for Quantum Advantage (C2QA) under Contract No.DE-SC0012704 (S.G., D.K., V.K.).
S.G. was supported in part by a Feodor Lynen Research fellowship of the Alexander von Humboldt foundation.
K.H. was supported by the National Natural Science Foundation of China (Grant Nos. 12275214, 12547107, 12247103, and 12047502), the Natural Science Basic Research Program of Shaanxi Province Grant Nos. 2021JCW-19, and Shaanxi Key Laboratory for Theoretical Physics Frontiers in China.
\appendix
\section{Lipatov's spin chain}\label{sec:framework2}

The remarkable connection between high-energy QCD and integrable quantum systems was established through Lipatov's mapping of the BFKL equation to an exactly solvable spin chain model~\cite{Lipatov:1993yb,Faddeev:1994zg}. The multicolor QCD Hamiltonian for $L$ reggeized gluons with nearest-neighbor interactions is expressed as
\begin{equation}
H_L = \sum_{k=1}^L H_{k,k+1},
\end{equation}
subject to periodic boundary conditions $H_{L,L+1} = H_{L,1}$. The local interaction Hamiltonian takes the form
\begin{align}
H_{j,k} &\!= P_j^{-1}\ln(z_{jk})P_j\! +\! P_k^{-1}\ln(z_{jk})P_k\! +\! \ln(P_j P_k)\! +\! 2\gamma_E \nonumber\\&= 2\ln(z_{jk}) + (z_{jk})\ln(P_j P_k)(z_{jk})^{-1} + 2\gamma_E,
\label{holomorphic-Hamiltonian}\end{align}
where $z_{jk} = z_j - z_k$ and $\gamma_E$ denotes the Euler constant. Here $z_j$ is the holomorphic transverse coordinate, and $P_j = i\,\partial/\partial z_j \equiv i\partial_j$ is the corresponding momentum operator (with $i$ the imaginary unit).

The system exhibits $SU(2)$ symmetry through the generators
\begin{equation}
S_k^+ = z_k^2\partial_k - 2s\,z_k, \quad S_k^- = -\partial_k, \quad S_k^z = z_k\partial_k - s,\label{eq:holosu2}
\end{equation}
for $k = 1,\ldots,L$. This construction establishes a correspondence between deep inelastic scattering processes and specific configurations of the quantum spin chain. The integrability of the resulting model allows for exact solutions using Bethe Ansatz methods, thereby providing analytical access to the eigenvalue problem. The integrability of Lipatov's spin chain is established through the existence of the fundamental $R$-matrix $R_{jk}^{(s,s)}(\lambda)$ which satisfies the Yang-Baxter equation
\begin{equation}
R_{jk}^{(s,s)}(\lambda) = f(s,\lambda) \frac{\Gamma(i{\lambda} - 2s )\Gamma(i{\lambda} + 2s + 1)}{\Gamma(i{\lambda} - J_{jk} )\Gamma(i{\lambda} + J_{jk} + 1)},
\end{equation}
where the complex-valued function $f(s,\lambda)$ normalizes the $R$-matrix. Furthermore, $\lambda$ is the spectral parameter. The superscript $(s,s)$ indicates that both the auxiliary and quantum spaces carry spin-$s$ representations. The operator $J_{jk}$ acting on the tensor product space $V \otimes V$, is defined through the operator identity
\begin{equation}
J_{jk}(J_{jk} + 1) = 2 \vec{S}_j \otimes \vec{S}_k + 2s(s + 1).\label{eq:op}
\end{equation}
Since all elements in this equation mutually commute, Vieta's formula can be consistently applied to solve it.
For the model with spin $s = 0$, which describes the nearest-neighbor interactions of reggeized gluons, the local Hamiltonian can be obtained from the fundamental $R$-matrix
\begin{equation}
H_{jk} = -\frac{1}{i} \frac{d}{d\lambda} \ln R_{jk}^{(s=0)}(\lambda)\bigg|_{\lambda=0},
\end{equation}
which yields the explicit form that recovers the BFKL local Hamiltonian \eqref{holomorphic-Hamiltonian}
\begin{equation}
H_{jk} = \psi(-J_{jk}) + \psi(J_{jk} + 1) - 2\psi(1),
\end{equation}
where $\psi(z)$ is the digamma function $\psi(z) = \Gamma'(z)/\Gamma(z)$, and $\psi(1)=-\gamma_E$. We adopt the shorthand notation $H_{jk} \equiv H_{j,k}$. The parameter $s(s+1)$ in equation~\eqref{eq:op} represents the Casimir eigenvalue of the ${SU}(2)$ representation, which vanishes for spin $s=0$ and $s=-1$.

The operator $J_{jk}$ is a solution of the equation \eqref{eq:op} when $s=0$ and $s=-1$, say $J_{jk}(J_{jk} + 1) = -(z_j - z_k)^2 \partial_j \partial_k$, where the right-hand side can be expressed in terms of the differential operators acting on the holomorphic coordinates.
This is a description of DIS in QCD \eqref{holomorphic-Hamiltonian} by a spin chain with $s = 0$.

However, the spin $s = 0$ model does not admit a nontrivial highest weight (vacuum) state , which obstructs the direct application of the Bethe Ansatz. By performing an appropriate similarity transformation to the standard quantum mechanical representation, the spin $s = 0$ model can be mapped to the spin $s =-1$ XXX Heisenberg chain, which can be solved exactly using the algebraic Bethe Ansatz method~\cite{Tarasov:1983cj}. The Bethe equations for the eigenvalues take the standard form, and the resulting energy eigenvalues are real and symmetric. This establishes a precise correspondence between high-energy QCD and the integrable spin system.

The fundamental monodromy matrix is constructed as the ordered product of fundamental Lax operators $L^{(s,s)}_{f,k}(\lambda)=R^{(s,s)}_{f,k}(\lambda)$ along the lattice ~\cite{Tarasov:1983cj,korepin1993quantuminversescatteringmethod},
\begin{equation}
T_f(\lambda) = L_{f,L}^{(s,s)}(\lambda)L_{f, L-1}^{(s,s)}(\lambda) \cdots L_{f,1}^{(s,s)}(\lambda),
\end{equation}
where each fundamental Lax operator acts simultaneously on the auxiliary and the quantum spin-$s$ spaces.
The associated fundamental transfer matrix is then defined as the trace of the monodromy matrix over the auxiliary space,
\begin{eqnarray}
\tau(\lambda)=\mbox{tr}_f\, T_f(\lambda),\quad
[\tau(\lambda),\tau(\mu)]=0,
\label{eq:fundamental-transfer}
\end{eqnarray}
which shows that transfer matrices commute for different values of the spectral parameter.

The commutativity relation of the fundamental transfer matrix for different values of the spectral parameter ensures the existence of an infinite number of conserved quantities, establishing the complete integrability of the model. This infinite set of conservation laws is generated by expanding the transfer matrix in powers of the spectral parameter, with the Hamiltonian emerging as the first non-trivial charge in this hierarchy.

The explicit construction of conserved quantities proceeds systematically. In particular, both the Hamiltonian of $s = -1$ and spin $s = 0$ models can be obtained from the first-order derivative of the transfer matrix:
\begin{align}
H_L^{(s=-1)} &= -\frac{1}{i}\frac{d}{d\lambda}\ln \tau^{(s=-1)}(\lambda)\Big|_{\lambda=0}, \\
H_L^{(s=0)} &= -\frac{1}{i}\frac{d}{d\lambda}\ln \tau^{(s=0)}(\lambda)\Big|_{\lambda=0}. 
\end{align}

The one-to-one correspondence between the XXX spin chains with $s = -1$ and spin $s = 0$ may be described by a similarity transformation which is based on the relation between the Lax operators and the definition of the transfer matrix \eqref{eq:fundamental-transfer}. Hence, the local spin $s = 0$ Hamiltonian can be transformed to the spin $s = -1$ Hamiltonian
\begin{equation}
H_L^{(s=-1)} = (z_{12}z_{13} \cdots z_{1L})^{-1}H_L^{(s=0)}z_{12}z_{13} \cdots z_{1L}.
\end{equation}
This means that the two Hamiltonians have identical eigenvalue spectra.
Despite this spectral equivalence, the two models differ in an essential way. The spin $s=-1$ chain admits a well-defined highest-weight (vacuum) state and can be solved exactly using the Bethe Ansatz, allowing for a transparent spectral analysis. In contrast, in the spin $s=0$ case the Bethe equations become degenerate (see Eq.~\eqref{eq:betheXXX}), and the absence of a nontrivial vacuum state obstructs a direct application of the Bethe Ansatz.
For this reason, in what follows we will primarily concentrate on the spin $s=-1$ model.

On the other hand, we define the monodromy matrix based on the $L$ operator with spin-$1\over2$ auxiliary space:
\begin{equation}
L_{a,k}^{({1\over2},s)}(\lambda) = \left(\begin{array}{cc}
\lambda \mathbb{1} + iS_k^z & iS_k^- \\
iS_k^+ & \lambda \mathbb{1} - iS_k^z
\end{array}\right),
\end{equation}
where the matrix elements are expressed in terms of the ${SU}(2)$ generators acting on the $k$-th site of the quantum space of spin-$s$.

We further define the associated transfer matrix as
\begin{align}
t(\lambda) &= \text{tr}_a[L_{a,L}^{({1\over2},s)}(\lambda) \cdots L_{a,1}^{({1\over2},s)}(\lambda)] \nonumber\\
&= \text{tr}_a \begin{pmatrix} A(\lambda) & B(\lambda) \\ C(\lambda) & D(\lambda) \end{pmatrix} \nonumber\\
&= A(\lambda) + D(\lambda), \label{eq:aux-transfer}
\end{align}
where the trace is taken over the auxiliary space and the monodromy matrix elements encode the scattering properties.
And then we have
\begin{eqnarray}
[t(\lambda),t(\mu)]=0.
\end{eqnarray}
One can get a family of mutually commuting conservation laws of the model.

Both of the two transfer matrices $\tau(\lambda)$ and $t(\lambda)$ act on the full quantum space of the model and they commute with each other for different values of the spectral parameters. The fundamental transfer matrix $\tau(\lambda)$ contains the local integrals of motion, including the Hamiltonian of the model. At the same time, the families $t(\lambda)$ and $\tau(\mu)$ commute \cite{Tarasov:1983cj},
\begin{equation}
[t(\lambda), \tau(\mu)] = 0, 
\end{equation}
so that the operators $t(\lambda)$ are also integrals of the motion for the original Lipatov's spin chain model.
In addition, the operator $t(\lambda)$ allows one to construct their eigenstates by means of the Bethe ansatz.

In the following, we will show how to construct the Bethe sates for Lipatov's spin chain along this line. 
Recall the spin operators given in Eq.~\eqref{eq:holosu2}, one finds with $s=-1$ 
\begin{equation}
S_j^+|\omega_j\rangle = 0, \quad S_j^z|\omega_j\rangle = -|\omega_j\rangle
\end{equation}
the solution $|\omega_j\rangle = 1/z_j^2$. Hence, the pseudovacuum is
\begin{equation}
|\Omega\rangle = (z_1^2z_2^2 \cdots z_L^2)^{-1}. 
\end{equation}

The Bethe states for spin $s = -1$ are then generated using the operator $B$ from Eq.\eqref{eq:aux-transfer}:
\begin{equation}
|\Psi(\{\lambda\})\rangle = B(\lambda_1)B(\lambda_2) \cdots B(\lambda_N)(z_1^2z_2^2 \cdots z_L^2)^{-1}. 
\end{equation}

These constitute the eigenvectors of Lipatov's spin chain. The eigenvalue of the transfer matrix $t(\lambda)$ as a function of spectral parameter $\lambda$ takes the factorized form:
\begin{equation}
(\lambda - i)^L \frac{Q(\lambda - i)}{Q(\lambda)} + (\lambda + i)^L \frac{Q(\lambda + i)}{Q(\lambda)}, 
\end{equation}
with the auxiliary function $Q(\lambda)$ defined as:
\begin{equation}
Q(\lambda) = \prod_{k=1}^N (\lambda - \lambda_k). 
\end{equation}

The corresponding Bethe equation, determining the parameters $(\lambda_1, \ldots, \lambda_N)$, is:
\begin{equation}
\left(\frac{\lambda_k + i s}{\lambda_k - i s}\right)^L = \prod_{j \neq k}^N \frac{\lambda_k - \lambda_j + i}{\lambda_k - \lambda_j - i}, \label{eq:betheXXX}\end{equation}
with $k = 1, \ldots, N$. We will use it with the substitution $s = -1$.

These represent periodic boundary conditions. To construct elementary excitations, one must change to antiperiodic boundary conditions. All solutions $\lambda_k$ of the above Bethe equations are real numbers, ensuring that the quantum lattice state of the system possesses real eigenvalues.
The Bethe Ansatz solution provides the complete eigenvalue spectrum through the characteristic equations
\begin{equation}
\left(\frac{\lambda_\alpha - i}{\lambda_\alpha + i}\right)^L = \prod^N_{\beta \neq \alpha} \frac{\lambda_\alpha - \lambda_\beta + i}{\lambda_\alpha - \lambda_\beta - i},
\label{eq:BAE}
\end{equation}
where $\{\lambda_\alpha\}$ are the Bethe roots that parameterize the eigenstates. The energy eigenvalues are expressed in terms of these roots, where $N$ is the number of quasi-excitations in the system. This exact solution enables the computation of correlation functions and provides the foundation for understanding the analytic structure of high-energy QCD scattering amplitudes through the correspondence with integrable quantum field theory.
The explicit expressions for eigenvalues of integrals of motion for arbitrary spin $s$ have been determined using algebraic Bethe Ansatz methods. For $s = -1$, the eigenvalues of the Hamiltonian are:
\begin{equation}
E = \sum_{j=1}^N \frac{-1}{i} \frac{d}{d\lambda_j} \ln \frac{\lambda_j + i}{\lambda_j - i} = \sum_{j=1}^N \frac{2}{\lambda_j^2 + 1},
\end{equation}
where $\{\lambda_j\}$ satisfy the Bethe equations \eqref{eq:BAE} for a fixed number of Reggeized gluons $L$. This relation yields the complete spectrum of the original holomorphic QCD model with Hamiltonian $H_L$.

\section{Quantum lattice nonlinear Schr\"odinger model}
\label{sec:framework3}
We begin with a brief description of the quantum lattice nonlinear Schr\"odinger model. The quantum lattice NLS equation was introduced as the quantum equivalent to the XXX spin chain with negative spin, representing a chain of harmonic oscillators. Let $\Psi_j^\dagger$ and $\Psi_j$ be the canonical creation and annihilation operators of the harmonic oscillator:
\begin{equation}
[\Psi_j, \Psi_k^\dagger] = \delta_{jk},\qquad \ \hat{n}_j = \left(1 + \frac{\kappa}{4}\Psi_j^\dagger\Psi_j\right)^{\frac{1}{2}},
\end{equation}
where $\delta_{jk}$ is the Kronecker delta function, $\kappa > 0$ is the coupling constant, and $\Delta > 0$ is a step of the lattice. The operators
\begin{align}
S_j^x &= \frac{i}{\sqrt{\kappa\Delta}}(\Psi_j^\dagger \hat{n}_j+\hat{n}_j\Psi_j),\\
S_j^y &= \frac{1}{\sqrt{\kappa\Delta}}(\hat{n}_j\Psi_j-\Psi_j^\dagger \hat{n}_j), \\
S_j^z &= -\left(\frac{2}{\kappa\Delta} + \Psi_j^\dagger\Psi_j\right), 
\end{align}
are the generators of an irreducible representation of $SU(2)$ algebra with a negative spin
$s =  - \frac{2}{\kappa\Delta}$. 

Even though the $SU(2)$ representation is infinite-dimensional in general, it can become finite-dimensional for special (negative) values of $\Delta$~\cite{Izergin:2009yc,Tarasov:1983cj}. Let us now elaborate the mapping between Bethe equations of the two models. First, the Bethe roots $\lambda_k$ of the quantum lattice NLS model obey the Bethe equations:
\begin{equation}
\left(\frac{1 + i\lambda_k\Delta/2}{1 - i\lambda_k\Delta/2}\right)^L = \prod_{j \neq k}^N \frac{\lambda_k - \lambda_j + i\kappa}{\lambda_k - \lambda_j - i\kappa}. 
\end{equation}
Comparison of the above modified Bethe equations and Bethe Eqs.~\eqref{eq:betheXXX} and~\eqref{eq:BAE} shows that they match for $\kappa = 1$ and $\Delta = 2$.

This means that the quantum lattice NLS model describes a more general XXX spin chain model with negative spin $s = -2/(\kappa\Delta)$. Moreover, it describes holomorphic QCD for $s = -1$, with $\Delta = 2$ and coupling constant $\kappa = 1$.

We can take the logarithm of the Bethe equations \eqref{eq:BAE} for the holomorphic NLS model (XXX spin $s = -1$) and define each number $n$ (integer or half-integer) as a vacancy. Some vacancies corresponding to Bethe roots are referred to as particles and the free vacancies as holes. The sum of the number of particles and holes is the number of vacancies. The density $\rho(\lambda)$ of vacancies in holes corresponds to:
\begin{equation}
2\pi\rho(\lambda) = \left(\int_{-\infty}^{-q}+\int^{\infty}_{q}\right)K(\lambda,\mu)\rho(\mu)d\mu + K(\lambda), 
\end{equation}
with
\begin{equation}
K(\lambda,\mu) = \frac{2}{1 + (\lambda - \mu)^2}, \quad K(\lambda) = K(\lambda,0). 
\end{equation}

\bibliography{main.bib}
\end{document}